# SIMPLICIAL QUANTUM GRAVITY ON A COMPUTER

S. Bilke, Z. Burda [1,2]

Fakultät für Physik Universität Bielefeld,
Postfach 10 01 31, Bielefeld 33501, Germany

J. Jurkiewicz [2]

The Niels Bohr Institute, Blegdemsvej 17,
DK-2100 Copenhagen Ø, Denmark

## Abstract

We describe a method of Monte–Carlo simulations of simplicial quantum gravity coupled to matter fields. We concentrate mainly on the problem of implementing effectively the random, dynamical triangulation and building in a detailed–balance condition into the elementary transformations of the triangulation. We propose a method of auto–tuning the parameters needed to balance simulations of the canonical ensemble. This method allows us to prepare a whole set of jobs and therefore is very useful in systematic determining the phase diagram in the two dimensional coupling space. It is of particular importance when the jobs are run on a parallel machine.

[1] A fellow of the Alexander von Humboldt Foundation.
[2] Permanent address: Institute of Physics, Jagellonian University, ul. Reymonta 4, PL-30 059, Kraków 16, Poland

# PROGRAM SUMMARY

*Title of program*:
GRAVZ2, AUTOTUNE

*Program obtainable from*:
bilke@physw.uni-bielefeld.de

*Computer*:
PARAGON XP/S 10, portable to other computers

*Operating system*:
OSF/1

*Programming language used*:
FORTRAN

*High speed storage required*:
Typical 400 Bytes / Simplex.

*Peripherals used*:
writes data and the last lattice configuration to disk. If required reads in old data and a configuration to continue a run from the disk. For a lattice with 4000 simplices the average size of a configuration file is 1.5 Mbyte.

*No. of lines in the program*:
6000 (together with supporting software).

*Nature of the physical problem*:
Four dimensional dynamically triangulated random surfaces as a regularization of quantum gravity coupled to matter fields.

*Method of solution*:
The program uses a standard dynamical Monte–Carlo scheme to produce configurations distributed according to the Gibbs measure of a discretized version of the Einstein Hilbert action with an additional term for a matter field. The geometry is updated by means of a set of elementary transformations basded on Alexander moves, ergodic in the space of triangulations with fixed topology. A geometrical update is performed alternating with a heat–bath for the matter sector.

*Typical running times*:
The time required for one update strongly depends on the coupling constants. Moreover it grows nonlinearly with a lattice size. In the critical region the time required to update the lattice including the matter sector is typically $150\mu s$ and $500\mu s$ per accepted move for a lattice with 4000 and 8000 simplices, respectively.
1

# 1. Introduction

The dynamical triangulation method has recently attracted great interest [1]-[12]. It was proposed as a regularization of gravity based on the discretization of a Euclidean path integral formulation [2]-[3]. In this method, the integration over the metric fields of the original continuum formulation is substituted by a sum over simplicial complexes, called simplicial manifolds or dynamical triangulations. The method was originally formulated, and very successfully used, in two dimensions [1], where the sum over the dynamical triangulations were shown to reproduce perfectly the measure of integrations over geometries. An additional feature of the method is that it permits to study non-perturbative problems. All this suggested to extend the method to higher dimensional gravity and ask the question about the existence of a continuum limit. In four dimensions a critical point of simplicial gravity was found. The transition is probably second order at this point. This opens a possibility to formulate a continuum theory independent of the regularization. At present it is not yet clear whether this transition corresponds to a nonperturbative point of quantum gravity. The main difficulty lies in the fact that the value of the average curvature in the regularized version does not scale to zero at the critical point. This implies that the curvature underlying theory blows up when the ultra violet cut-off approaches zero. It is believed, however, that by adding an appropriate term to the action one can shift the average curvature to zero. The curvature square terms proposed very long ago in the continuum approach, were also studied numerically for simplicial gravity [7],[8]. The most interesting region, namely the region of large couplings, seems however not to be available to Monte–Carlo simulations, because in this region the acceptance rate of the changes proposed by the algorithm is too low. Another way to generate nontrivial terms in a gravity action is to couple covariantly matter fields to it. By integrating matter out, one generates new terms in the effective gravity action. If the matter sector is critical, the interactions are long-ranged and they can contribute nontrivially to the effective gravity action modifying also the critical properties of the gravity sector. For example in the Ising model on a two dimensional random lattice the gravity entropy exponent $\gamma$ changes when the spin field becomes critical. The present paper is devoted to the basic features of the algorithm and the implementation used for simulating models of this type. The results and detailed description of the models themselves are presented elsewhere [8]-[10].

## 2. The model

The simplicial manifold (or triangulation) is constructed from equilateral 4d simplices, which we will call pentahedra, by identifying pairs of neighboring pentahedra which share a four dimensional face. A pentahedron has 5 vertices, 10 links, 10 triangles and 5 tetrahedra. The nearest neighborhood of a point on the triangulation, formed by pentahedra meeting there, has the topology of a 4d ball, which means that the triangulation is locally homomorphic to $R^4$. This imposes some relations between the numbers of simplices on the triangulation, known in the general case as the Dehn–Sommerville relations. In four dimensions they have the form :

$$5N_4 = 2N_3, \qquad 5N_4 - 4N_3 + 3N_2 - 2N_1 = 0 \qquad (1)$$

where $N_4, N_3, N_2, N_1, N_0$ denote the numbers of pentahedra, tetrahedra, triangles, links and points of the triangulation, respectively. Additionally the numbers of (sub)simplices are related by the Euler formula :

$$N_4 - N_3 + N_2 - N_1 + N_0 = \chi \qquad (2)$$

where $\chi$ is the Euler characteristic of the manifold. Altogether, for manifolds with fixed topology, and this is the case in our considerations, these three relations between the five numbers $N_i$ leave two of them, say $N_4$ and $N_2$, independent. They completely specify the size of a simplicial manifold. The first one, $N_4$, is the 4d volume of a manifold, the other one, $N_2$, is the total deficit angle of triangles on the manifold. The average curvature : $\langle R \rangle \propto N_2/N_4 - 2.097..$. A collection of manifolds with a fixed number of (sub)simplices is called the micro–canonical ensemble. Denote the number of states (triangulations) by $\mathcal{N}(N_2, N_4)$. Summing over all possible $N_2$ leads to the canonical ensemble with the 4d volume, $N_4$, fixed :

$$\mathcal{Z}(\kappa_2, N_4) = \sum_T e^{\kappa_2 N_2(T)} = \sum_{N_2} \mathcal{N}(N_2, N_4) e^{\kappa_2 N_2} \qquad (3)$$

The weight for a triangulation $T$, namely $k_2 N_2(T)$, is naturally provided by a discrete version of the Einstein–Hilbert term, that is for a fixed volume proportional to $N_2$. It is also possible to consider the grand-canonical ensemble by letting the volume fluctuate controlled by the cosmological term :

$$Z(\kappa_2, \kappa_4) = \sum_{T \in \mathcal{T}} e^{-\kappa_4 N_4 + \kappa_2 N_2} = \sum_{N_2} \mathcal{N}(N_2, N_4) e^{-\kappa_4 N_4 + \kappa_2 N_2} \qquad (4)$$

All three ensembles are equivalent to each other, in the sense that they are related by Legendre transforms. This observation can, however, hardly be used numerically, because to make the transformation one should know the behavior of the partition function in the whole range of parameters. Therefore, beforehand one should decide which ensemble to simulate. Most of the physically interesting questions are simple to formulate in the canonical ensemble. By means of finite size analysis one can study signals related to the phase transition and determine its order. In the frame of the canonical simulations one can also try to extract some information about the number of states $\mathcal{N}(N_2, N_4)$ from the baby universes distribution or the sum rules. In this respect it gives some insight into the behavior of $\mathcal{N}(N_2, N_4)$, while for microcanonical simulations it is not possible to compare numbers of states for different $N_2$ and $N_4$.

To discuss the matter sector let us come back to the microcanonical ensemble. The number of states $\mathcal{N}$ depends only on $N_2$ and $N_4$. One can, however, imagine that on a surface there are some excitations which depend on some invariant characteristic of the triangulation which is neither $N_4$ nor $N_2$. As an example consider the average of a certain power, $\alpha$, of the order of a triangle $\langle o(t)^\alpha \rangle$. By the order of subsimplex we mean the number of simplices, which share this subsimplex. One can control these excitations by introducing an additional coupling to the action. In fact, some of such couplings are very well motivated since they correspond to the higher derivative action used in the continuum formulation of 4d gravity. With the new term in the action, the weight of a triangulation explicitly depends on $\langle o(t)^\alpha \rangle$, or say generally on a certain characteristic, $c_T$, of the triangulation $\mathcal{N}(N_2, N_4, c_T)$. It is believed that this kind of coupling can cure the problems encountered in formulating continuum limit of simplicial gravity. One can introduce the coupling to the model by hand, as it was done in case of curvature square term in [7], or dynamically by introducing a new field covariantly coupled to gravity. If one integrates out the matter field one gets an effective weight for a triangulation $T$ :

$$\mathcal{N}_{eff}(N_2, N_4, c_T) = \sum_{\text{matter on } T} e^{S(\text{matter})} \tag{5}$$

A few models of this type have been already studied [8]-[10]. Because the presented algorithm is very general and can be used to any matter field with a local action the following description of the algorithm does not assume any specific form of the action for the matter sector. To be specific, while presenting implementation, we will, however, refer to the $Z_2$ gauge model [10], with the action $S(\beta) = -\beta \sum_{t \in T} o(t) \{\sigma\sigma\sigma\}_t$, where $o(t)$ is the order of a triangle $t$ and $\sigma\sigma\sigma$ is the product of three $Z_2$ link variables lying on the edges of $t$. The order of a triangle, $o(t)$ plays the role of a two dimensional volume dual to the triangle, which multiplied by the area of the triangle gives the 4d volume the plaquette $\{\sigma\sigma\sigma\}_t$ is associated to. This factor $o(t)$ assures that the matter field is coupled covariantly to gravity [13]. An important feature of this action is that it couples the matter field directly to the local manifold curvature, represented by the order of a triangle. A closer motivation to study this kind of models and the results are presented in [8]-[10] .

## 3. The method of simulation

Simplicial quantum gravity is simulated by means of a Markov chain in the space of four dimensional simplicial manifolds. The chain has a stationary distribution equal to the Gibbs measure defined in (3). This is achieved by requiring that the chain is ergodic and that he Markov probabilities fulfill a detailed–balance condition. A set of local, topology preserving transformations ergodic in the grand canonical ensemble is known for a very long time. They are called the Alexander moves [15]. Though they are ergodic, they cannot be directly applied to Monte Carlo simulations in an efficient way because in the physically interesting region of coupling constants they have low acceptance [11]. Therefore the Markov chain looses its mobility in the space of triangulations. Another set of more practical moves (transformations), which turned out to have higher acceptance rate, was suggested in [12]. In four dimensions there are five different moves in this set. They are defined as follows. Let us enumerate the moves by the index $i$ which runs from 0 to 4. The $i$-th move substitutes a $i$-dimensional simplex shared by $(5-i)$ pentahedra by a $(5-i)$-dimensional simplex shared by $i$ pentahedra, in such a way that the pentahedra after the move have the same $3d$-boundary as the original ones had before. It is convenient to represent the moves schematically. The move 0 substitutes a point (0 dimensional simplex), denoted by 6, shared by 5 pentahedra on the left hand side of (6) by one pentahedron :

$$123\underline{6} + 1235\underline{6} + 1245\underline{6} + 1345\underline{6} + 2345\underline{6} \leftrightarrow \underline{12345} \qquad (6)$$

where 1, 2, 3, 4 and 5 stand for different points. When read from the right to left, the scheme (6) represents move 4. Notice that the $3d$-boundary of pentahedra is the same before and after the move and consists of the $3d$-simplices : 1234, 1235, 1245, 1345 and 2345.

Similarly the set of the moves 1,3 is :

$$\underline{12}345 + \underline{12}346 + \underline{12}356 + \underline{12}456 \leftrightarrow 1\underline{3456} + 2\underline{3456} \qquad (7)$$

The move 1 starts with the link 12 common to four pentahedra and ends up with the tetrahedron 3456 common to two pentahedra. The move 3 does the opposite.

The move 2 is self–dual :

$$12\underline{456} + 13\underline{456} + 23\underline{456} \leftrightarrow \underline{123}45 + \underline{123}46 + \underline{123}56 \qquad (8)$$

and it flips the triangle common for three pentahedra.

In all moves six points and six pentahedra are involved. One can easily check that they always form the 4d surface of a 5d sphere. Therefore, each move can be treated as substitution of a part of the sphere (found on a simplicial manifold) by its supplement. This assures that a topology of simplicial manifold is preserved by the move. Because both sets of simplices of the triangulation have the same boundary, a substitution is always possible. But not all changes of the triangulation proposed by a move can be accepted. The requirement that a triangulation is locally homomorphic to $R^4$, imposes the restriction to accept only those moves, which create only essentially new (sub)simplices, not yet present on the triangulation. For example, a move that would produce a link between vertices already joined by another link is rejected. More generally, a move which would lead to any double connection on a lattice is rejected.

The set of moves described above is equivalent to the Alexander moves in the sense that each move can be obtained by a certain sequence of the Alexander moves, and vice versa. This proves their ergodicity in the grand–canonical ensemble.

Let us now describe more precisely how to build in a detailed–balance condition into the Markov probabilities. In some respects the situation differs from that in standard MC simulations of field–theoretical models on a regular lattice.

Firstly, the proposals for each pair of mutually inverse moves 0,4 (eq. 6) or 1,3 (eq. 7) are not the same in both directions. To see the consequence of this, suppose that for a certain configuration $A$ we want to perform a move $i$ and balance it with the inverse transition using move $(4-i)$, from the configuration $B$ obtained from $A$. Denote the transition probabilities by $\mathcal{P}(A \to B)$ and $\mathcal{P}(B \to A)$, respectively. A transition from $A$ to $B$ is realized by two following steps. First, we pick up a $i$–dimensional simplex shared by $(5-i)$ 4d simplices on $A$ with a probability $1/n_i(A)$, and afterwards, at this particular simplex, we perform move $i$ with a move probability, $P_i(A \to B)$. This probability is related to the transition probability by : $\mathcal{P}(A \to B) = 1/n_i(A)\, P_i(A \to B)$. The detailed balance condition for the move probability reads :

$$\frac{e^{-S(A)}}{n_i(A)} P_i(A \to B) = \frac{e^{-S(B)}}{n_{4-i}(B)} P_{4-i}(B \to A), \quad i = 0, \ldots 4. \qquad (9)$$

which in fact is three independent sets of equations. We would like to emphasize the difference between the move probabilities $P_i$ which is a probability of performing an operation (6–8) at a specific place on a lattice and the transition probabilities. This difference is usually absent in standard MC simulations of systems on fixed lattices since the balanced transitions are symmetric and the combinatorial prefactor of the type $1/n_i$ is the same on both the sides of a detailed–balance equation. Note, that for the self dual canonical move (8) the prefactor can be dropped in (9), because a move 2 does not change the number $n_2$. If one wants to associate a physical meaning to the combinatorial prefactors $1/n_i$ in the detailed balance condition, one can say that they reflect a purely geometrical change of the number of states between different canonical ensembles and they enter the equations (9) to balance the entropy change coming from the integration measure of the geometry sector.

The changes are grand–canonical, with the consequence that the number of degrees of freedom for the matter field varies. When performing grand–canonical moves (6-8) one has to create or erase matter field variables. The detailed balance equations leave a freedom how for a given configuration a new one can be proposed and how to choose the move probabilities. The simplest way of fulfilling the detailed balance condition (9) is to assign new field variables at random with a uniform distribution on the cartesian product of symmetry groups for each created matter degree of freedom and accept a change with the Metropolis probability. The drawback of this procedure is that as a consequence of random choice of fields the acceptance rate is very low, especially for moves 0,4 (eq. 6), where ten new links are created or deleted simultaneously. Significant improvement can be achieved by proposing new fields with a probability dictated by the Gibbs measure :

$$\pi_i(s_B) = \frac{e^{-S(s_B)}}{\sum_{s_B} e^{-S(s_B)}} \qquad (10)$$

where $s_B$ is a field configuration that can be obtained from $A$ by performing move $i$ on a specific place on the lattice. The sum in the denominator runs over all these configurations. The main difficulty to apply directly a heat bath procedure is that the analogous denominator for the probability of configurations $s_A$ that can be obtained by the inverse move in the same place of the lattice : $\pi_{(4-i)}(s_A) = e^{-S(s_A)}/\sum_{s_A} e^{-S(s_A)}$ differs from that for $s_B$, because moves $i$ and $(4-i)$ create different sublattices and matter field subspaces. Therefore $\pi$'s cannot directly be used as heat–bath probabilities. We propose to split the transition into two steps. The first step is to accept or reject a change of geometry regardless of the values of the new matter fields on the new sublattice created in a move. Therefore its probability does not depend on the new field configuration on the new sublattice created by the move, but only on the shape of this new sublattice, and on the fields in the vicinity of the place where move is performed. We denote the probability of this step by $p_i(t_A \to t_B)$. The letter $t$ (for triangulation) is used to emphasize that this probability only depends on the subtriangulation and not on the fields on it. After a change of a triangulation $t_A \to t_B$ is accepted, new fields according to the distribution $\pi_i$ are assigned. Altogether we can write the probability of this move as $P(A \to B) = p(t_A \to t_B)\pi_i(s_B)$. Inserting this into the detailed balance condition (9) and summing both sides over $s_A$ and $s_B$ one gets the equation :

$$\frac{\sum_{s_A} e^{-S(s_A)}}{n_i(A)} p_i(t_A \to t_B) = \frac{\sum_{s_B} e^{-S(s_B)}}{n_{4-i}(B)} p_{4-i}(t_B \to t_A), \tag{11}$$

which is fulfilled by :

$$p_i(t_A \to t_B) = \min\left\{1, \frac{n_i(t_A)}{n_{4-i}(t_B)} \frac{\sum_{s_B} e^{-S(s_B)}}{\sum_{s_A} e^{-S(s_A)}}\right\} \quad i = 0, \ldots, 4. \tag{12}$$

The last equation is one of many solutions of (11). It essentially means that the change of triangulation is done by a Metropolis algorithm that includes besides the factor $\kappa_2 \Delta N_2 - \kappa_4 \Delta N_4$ coming from the change of geometrical part of the action also the ratio of the volumes of the field configurations created by the moves.

So far we have described the grand canonical algorithm. In principle, it can also be used in canonical simulations in the following way. The grand canonical algorithm generates, among different volumes, samples of a certain volume $N_4^0$. These samples can be used in simulations of a canonical ensemble for the volume $N_4^0$. Because the volume of the canonical system is fixed, the coupling $\kappa_4$ is a free variable and can be tuned to make the distribution of volumes concentrated around $N_4^0$. Unfortunately, this distribution is very broad, and therefore the algorithm spends most of the time on volumes different from $N_4^0$. Moreover, the volume fluctuations are sometimes too large for the size limits imposed by the computer implementation. The situation can be improved by adding to the action a potential that controls volume fluctuations [14]. The simplest potential of this type has the form : $V(N_4) = \delta|N_4 - N_4^0|^\alpha$. A typical choice for $\alpha$ is 1 or 2. For $N_4 = N_4^0$, the potential vanishes, so it does not change the action for the canonical ensemble. By tuning $\delta$ one can suppress the volume fluctuations, and make the volume distribution narrower. The canonical ensemble is not affected by the shape of this distribution. But for too large values of $\delta$ this can spoil ergodicity. In particular, for the limiting case of very large $\delta$, the volume fluctuations are completely suppressed and only moves 3 (eq. 8) can be executed. The value of $\delta$ should be chosen to balance between the two conflicts coming from the mobility (ergodicity) of the algorithm and the narrow volume distribution. We found experimentally that below a certain value of $\delta$, which is of order of 0.2 for $\alpha = 1$, measurements of physical quantities do not depend on this value. Further lowering of $\delta$ changes only the range of fluctuations in the volume, the acceptance rate and the autocorrelation times.

The potential modifies the transition probability $p(t_A \to t_B)$ by an additional prefactor $e^{V(N_{4,A}) - V(N_{4,B})}$. In the case of $\alpha = 1$, which we use in our simulations, it is equal to $e^{\pm \delta \Delta N_4}$ above or below $N_4^0$, respectively, which alternatively means that we change the coupling $\kappa_4$ to be $\kappa_4 \pm \delta$ for volumes greater or smaller than $N_4^0$.

## 4. Implementation

While implementing a simplicial manifold on a computer, one should take two things into account. An implementation must uniquely specify the manifold, and it should give the possibility to reconstruct basic information about the neighborhood of each (sub)simplex. Among different implementations one can imagine two extreme ones, which we call minimal and maximal coding. In the former case one keeps only minimal information, namely for each pentahedron on the lattice one keeps addresses of the 5 neighboring pentahedra and the 5 points at its vertices. In case of the maximal coding one keeps for each (sub)simplex the number and the addresses of *all* (sub)simplices emerging from it. The main difference between these two approaches becomes clear, if one considers in more detail the elementary moves. In the first part of a move one has to pick up a (sub)simplex on the lattice, look for a double connection and then determine the action needed to evaluate the Metropolis probability. This part is done much faster in the maximal coding because the information about neighboring simplices is easily available. In the minimal coding one has to recover this information by going many times through the list of all simplices on the lattice to find out if a given (sub)simplex specified by points, is a (sub)simplex on the lattice. This procedure is very time consuming.

The second part of a move is to apply the geometrical update, unless it was rejected in the first part. This part is much easier in the minimal coding because the number of objects to update is much less than for the maximal one. The first part of a move is executed much more frequently than the second one because many of attempted moves are rejected either by the double connection test or by the Metropolis question. Therefore in general the maximal coding is faster. The difference in speed is especially pronounced when the acceptance rate is low. This favors maximal coding. On the other hand, memory limitations favor the minimal coding, because one can code larger lattices.

## 4a. Local structure

None of this implementations is used in practice. We mention them to show the problem of balancing between speed and memory and to give a kind of frame of reference, which helps to place a specific implementation, by saying that it is closer to the minimal or maximal coding. Our implementation is closer to the maximal coding. We code pentahedra maximally, namely we hold for a pentahedron addresses to 5 points at its vertices, 10 links at its edges, 10 triangles at its two dimensional faces, and the five neighboring pentahedra. The maximally coded pentahedra form the basis for the implementation. If, for a (sub)simplex on the lattice, we want to get the addresses of the neighboring (sub)simplices we first refer to a pentahedron and then through it to the addresses of each (sub)simplex. In other words, the philosophy of accessing addresses of neighbors is based on the pentahedron bridge : **(sub)simplex → pentahedron → (sub)simplex**. To facilitate referring from a (sub)simplex to a pentahedron in the first part of the bridge we introduced in the code matrices of pointers from (sub)simplices to pentahedra. The number of pentahedra to which a (sub)simplex belongs, changes during the run. To avoid holding a dynamical list of all pentahedra attached to the (sub)simplex, we keep the address of only one of them. This suffices because the other nearest pentahedra can be sequentially found by using addresses of neighbors kept in the pentahedra themselves. In fact, we use the pentahedron bridge mainly in the first part of the moves (eq. 6–8), where the nearest neighborhood of a (sub)simplex is well defined, and therefore can be easily reconstructed from addresses kept in the pentahedra. The sequential procedure of finding neighboring pentahedra is more time consuming than just taking the pentahedra

```
      parameter(       nmax=3000
     &,                nsmax=4*nmax+2,ntmax=10*nmax+10,nlmax=5*nmax+10
     &,                npmax=nmax+5,nt3max=3*ntmax,nt3lmax=nt3max+nlmax
     &,                nlpmax=nlmax+npmax,nt3max1=nt3max+1
     &,                nlmax1=nlmax+1)
      common /lattice/ sn(5,nsmax),st(10,nsmax),sl(10,nsmax),sp(5,nsmax)
     &,                ts(ntmax) ,ls(nlmax) ,ps(npmax)
     &,                tcn(ntmax),lcn(nlmax),pcn(npmax)
     &,                tf(0:nt3lmax),tb(0:nt3max),tp(nt3max),tl(nt3max)
     &,                lf(0:nlpmax),lb(0:nlmax),lp(nlmax),spin(nlmax)
      dimension        lt(nlmax),pl(npmax),s(5)
      equivalence      (tf(nt3max1),lt(1))
      equivalence      (lf(nlmax1) ,pl(1))
```

Table 1: Variables used to describe the random lattice

from prepared lists, but the time is paid back because one does not have to update all the lists for the many (sub)simplices involved in a move. The next advantage is that one avoids an additional storage of the information which is rarely used. In the program, the data structure needed for the pentahedron bridge is represented by the first two lines of the common block `lattice` declaration shown in the table 1. As a naming convention, we use the letters <u>s</u>implex, <u>t</u>riangle, <u>l</u>ink and <u>p</u>oint to distinguish the different kinds of (sub)simplices. We used the word <u>s</u>implex for <u>p</u>entahedron to avoid the conflict with <u>p</u>oint. The letters are combined in the identifiers which have always the two–piece structure `xy` meaning that an object `x` points to `y`. For example, the name `st` means that 4d <u>s</u>implex points to <u>t</u>riangle. The matrix `st(10,n)` holds pointers to the 10 sub-triangles of simplex `n`, or `ts(n)` being a matrix of pointers from triangles to simplices, contains a pointer to one simplex, triangle `n` is part of. The third line of the common block `lattice` contains vectors with orders of triangles, `tcn`, links, `lcn`, and points, `pcn`. The ranges of the arrays are controlled by the <u>max</u>imal <u>n</u>umbers of 4d <u>s</u>implecies, `nsmax`, <u>t</u>riangles, `ntmax`, <u>l</u>inks, `nlmax` and <u>p</u>oints, `npmax`.

## 4b. Double connections

A direct use of the pentahedron bridge in the test for double connection would be more complicated than in identifying the (sub)simplices engaged in a move, because in this case the pentahedra, to which a (sub)simplex belongs form unknown structures, and the number of them can be very large.

As an example consider move 3 which creates a link between the points **1** and **2** while going from the right to the left hand side of the equation (7). Before accepting the move we have to check if the link **12** already exist on the lattice. If one does it by use of the pentahedron bridge, one has to find all the pentahedra connected to one of those two points, and then check ten links, for each of them. Both steps are quite time consuming. We found it is much more efficient to supplement the pentahedron bridge by the **point → link → point** bridge which facilitates the access to neighboring points. Because the number of links emerging from a point changes during the run, we have to construct a dynamic structure which allows us easily to insert and remove links. Denote addresses of points in the bridge by `p1` and `p2` and assume that `p1>p2`. A link, which joins two points in the bridge, is associated with the one which has the larger address, `p1`. Each of these points may belong to many bridges simultaneously. The pointer `pl(p)` holds the address of one link, `l`, emerging from the point `p`. If no link is associated with a point `p`, entry `pl(p)` is set to zero. The other associated links are referenced with the help of a chained list, constructed by means of arrays `lf(l)` which points (**f**orward) from a current link `l` to the next link associated with the same point and `lb(l)` which points (**b**ackwards) to the previous link in the chain. The `lf` chain is zero-terminated, in `lb` the last link in the chain references to the *point* the link is associated to. The bridge is completed by the variable `lp(p)` which gives the address, `p2`, of the second point at the end of the link `l`. To check for an existing link between `p1`, `p2` we have to visit all links in the chain starting at `pl(p1)` and see if there is one entry `l` with `lp(l)=p2`. The update of the list of links emerging from a point `p` is done as follows. A new link in the chain is always inserted at the origin of the chain. This is done by setting `lb(pl(p))=l, lf(l)=pl(p), pl(p)=l`. To remove a link `l` from the chain, we set a new bond between its neighbors `lf(l), lb(l)` by `lf(lb(l))=lf(l)` and `lb(lf(l))=lb(l)`, omitting in this way `l` in the chain. A problem arises for the first entry in the chain. Then `lb` refers to a point and not to a link. We do not have to check for this situation, when we additionally use the following memory layout:

    equivalence     (lf(nlmax1) ,pl(1))    .

The ranges of `lf` and `pl` are glued one beyond the other. If we now denote a reference to a point `p` in the `lb` chain by `p+nlmax` we automatically have `lf(nlmax+p)=pl(p)` what we want.

In move 2 (eq. 8) one has to avoid a double triangle. We support the detection of such situation with the help of the **link → triangle → point** bridge. In order to have access to all triangles which are connected to a link each triangle `t123` is a member of three independent chains containing three rotated copies of triangle `t123`. We keep also an inverse matrix leading from each rotated copy of the triangle back to the link, `tl`. The purpose of this is to have fast access to link variables while computing staples needed to update links in gauge field models. The arrays `tf`, `tb`, `tp` (analogous to `lf`, `lb`, `lp` in point-link-point bridge) represent this structure in the implementation. The memory layout and one chain of triangles is shown in the figure 1.

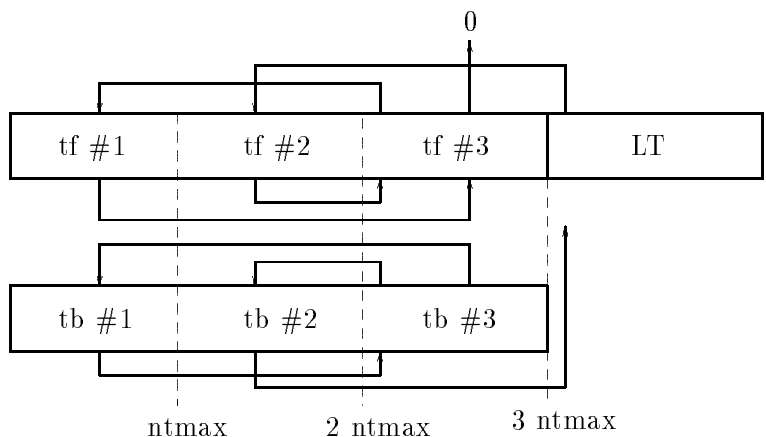

Figure 1: The memory layout for the link → triangle → point bridge. There are three independent copies for each triangle.

To complete the discussion about the tests for a double connection, we should also describe how to check for an existing tetrahedron **3456** in the move 1 (eq. 7). We postpone this to the section **4d.** when we have described the rotation of a pentahedron which facilitates the pentahedron bridge used in this test. Now we only mention that the idea is to fix triangle `t345` and visit all neighboring points to check if one of them is the point **6** needed to complete tetrahedron **3456**.

## 4c. Memory management

In the grand canonical simulations the size of the triangulation varies in a run. Some (sub)simplices are created on the lattice, some others are erased. Erasing a (sub)simplex leaves empty entries in the statically allocated matrices which define the lattice connections. To trace (sub)simplices as used/unused on the current triangulation, we introduce a kind of memory management in form of a sorted list of pointers to (sub)simplices. The list is mainly used to improve finding a candidate for each move $n$, namely a $n$-dimensional (sub)simplex of order $(5-n)$. The number of $n$-dimensional (sub)simplices of order $(5-n)$, is much less (for $n \leq 3$) than the number of all $n$-dimensional simplices used in the current triangulations, which itself is less than the number of statically allocated $n$-dimensional simplices. Therefore the probability of choosing a right candidate at random from the whole list is small. The memory management allows us to make this step in more efficient manner.

The main ingredient for memory control is a sorted list of $n$-dimensional (sub)-simplicies. They are stored in the matrices `is`, `it`, `il` and `ip` providing <u>i</u>ndexing of 4d <u>s</u>iplicies, <u>t</u>riangles, <u>l</u>inks and <u>p</u>oints. Let us first fix attention on the list of pentahedra `is`. The first `nsu` entries in this array contain vectors to storage cells currently used on the lattice. The remaining entries point to different unused storage cells. The variable `nsu` contains the number of simplices, $N_4$, currently used on the triangulation. To create a new simplex, we can use the storage cells pointed to by the first entry of type unused in the index-table: `is(nsu+1)`. When we now increase `nsu` by one this entry is marked as used. In the opposite case when we want to remove a simplex `s` from the lattice we can decrease `nsu` by one and thereby mark the last used entry as unused. To facilitate finding

```
 common /lists/    si(0:nsmax),is(0:nsmax),ti(0:ntmax),it(0:ntmax)
&,                 li(0:nlmax),il(0:nlmax),pi(0:npmax),ip(0:npmax)
 common /l_size/   nsu,ntu,nt3,nlu,nl4,npu,np5
```

Table 2: Arrays and supporting variables used for memory management

of the entry of a given simplex `s` we introduce the array `si(s)` giving us the position `i` in `is(i)` that controls simplex `s`. The correct entry is marked as unused when we now exchange the contents of `is(si(s))` and `is(nsu+1)`. The same method is applied for the other (sub)simplices. In these cases we can improve the access to a $i$-dimensional subsimplex of order $(5-i)$ required for the moves by introducing a new status for a storage cell. For example in case of triangles we need to know which triangles are of order 3. We sort the index table `it(i)` in the following way. The first `nt3` entries (`nt3` is the number of triangles of order 3) contain the address of triangles that are used and of order 3. The next entries up to `ntu` are used and of different order. The remaining entries are unused. The code to access a triangle of order 3 at random now is

```
t=it(min(int(ranmar()*nt3)+1,nt3))
```

where `ranmar()` is a random number on $[0,1]$.

## 4d. Pentahedron rotation

In the next step of a move one needs to know the information about neighbors, namely for $i$-dimensional (sub)simplex belonging to the $(5-i)$ pentahedra. In the moves, however, once one knows the address of one pentahedron, one can easily find the remaining ones, without referring to any list, because one knows that they are a part of the minimal $5d$ sphere, and therefore they form one of the structures in (eq. 6-8). In the easiest case, move 4, we have to insert a point into a simplex `s12345`. At the end of this procedure we will have 5 simplices. The internal structure, creating subsimplices, neighbor relations and so on is the same for all situations and it is hard-coded in the program. Additionally we have only to update the list of nearest–neighbors for the surrounding pentahedra.

For the other moves the situation is somewhat more complicated. To recover the local structure we impose some relations on the indices a, b used in the arrays st(a,n), sl(a,n), sp(b,n), sn(b,n). The point sp(b,n) is opposite to neighbor sn(b,n). Similarly for links and triangles : a link sl(a,n) and a triangle st(a,n), for the same index a, are built from complementary vertices of pentahedron n. A relation between between a and b is established with the help of a list:

```
 data ((lt_p(b,a),b=1,5),a=1,10)/  1,2,   3,4,5
&,                                 1,3,   2,4,5
 .........
&,                                 4,5,   1,2,3 /
```

which for example means that the link a= 1 connects the points b= 1, 2. The points b= 3, 4, 5 are not part of this link and therefore belong to triangle a= 1. In the coding of the program we use the naming convention $sp_1p_2p_3p_4p_5$ to identify a simplex consisting of points $p_1, \cdots, p_5$. Enumerating the indices does not lead to a new configuration but it can be interpreted as a rotation of a simplex. In the program this can be used to choose a convenient orientation.

To show how this construction works consider move 0, where we have to remove a point p6 from the lattice. First we have to identify the five inner simplices which are to be removed from the lattice. With the help of ps(p6) we can identify one of the 5 simplices that contains p6. Denote it by s23456. Once we know the index b such that p6=sp(b,s23456), we can easily find four other simplices having the point p6. They are namely the neighbors of s23456, sn(a,s23456) for a different from b. The destruction of the inner structure is hard-coded for the situation, where p6=sn(5,s23456). This is usually not the case and we use the freedom of orientation to rotate s23456 so that we encounter the hard-coded situation. This rotation and similar rotations for the other moves are performed with the help of some precomputed lists.

```
 common /orient/  i1_k5(5),i2_k5(5),i3_k5(5),i4_k5(5)
&,                i2_k1(5),i3_k1(5),i4_k1(5),i5_k1(5)
&,                i4_k123(5,5,5),i5_k123(5,5,5)
&,                i_k_all(14),l(5,5),lt_p(5,10)
```

In our case `i1_k5(b)` gives the first index of the rotated situation, `i2_k5(b)` the second index and so on. The other matrices are used in the other moves. The matrices `i4_k123` and `i5_k123` give positions of two points in a rotated pentahedron for three others known. The matrix `i_k_all` gives a position of one point if all others are known. The matrix `l(5,5)` enumerates links.

The table `lt_p` and the rules imposed on the indices `a`, `b` is also useful for the double connection check of move 1, where a link `l12` sharing 4 simplices is removed and replaced by a tetrahedron `3456`. In this case we have to check if this tetrahedron is already present on the lattice. The geometric idea behind the following steps is to walk around triangle `t345` and visit all simplices attached to it. If one of these contains the remaining point `p6` the move has to be rejected. We start the move by choosing a link `l12` of order 3. With the help of `ls(l12)=s12345` we get one of three pentahedra to which it belongs. It has two neighbors `s2345x`, `s1345x` opposite to points `p1`, `p2` which share a triangle `t345`. The naming convention is that a simplex `s2345x` has four points 2, 3, 4, 5 common to `s12345` and one different, symbolically denoted by `x`. One can identify these two pentahedra as `s1345x=sn(2,s12345)`, `s2345x=sn(1,s12345)`. The next step is to walk in the direction of `s1345x`. The simplex `s1345x` has again two neighbors sharing `t345`. They can be identified as follows. First we fix the index `b` which specifies a position of the triangle `t345` in the simplex `st(b, s1345x)=t345`. Then, by means of the earlier described table `lt_p` we can also recover the positions of these two neighbors `sn(lt_p(1,b),s1345x)`, `sn(lt_p(2,b),s1345x)`. One of this simplices lies in the direction we came from, so the other in the new direction. To continue our trip around `t345` we choose the pentahedron lying in the new direction and we repeat the whole procedure. The loop is terminated when the program encounters a simplex containing `p6` or when the walk is complete *ie* when we reach the second neighbor `s2345x` of the starting simplex `s12345`.

### 4e. Starting configuration

To set up a starting configuration for a run, it is sufficient to build by hand a minimal valid configuration and then use move 4 to increase the lattice size to the required value $N_4^0$. For a sphere the minimal valid configuration is the 4 dimensional surfaces of 5d simplex. It consists of six pentahedra glued in such a way that each two of them are neighbors. To save some work and not to fill by hand the tables for six pentahedra we start from the invalid configuration of only two pentahedra glued together along all their 3d faces. To one of them we apply move 4. It is safe operation because it acts only inside a pentahedron without referring to its neighborhood. After correcting nearest neighbors relations `sn` we get a well defined minimal sphere.

## 4f. Parameter tuning

To run a job simulating canonical–ensemble for given coupling constants $\beta, \kappa_2$ one needs to tune $\kappa_4$, which makes the system fluctuate around the desired canonical volume $N_4^0$. It can be done by hand by executing some updates on the lattice with a trial value for $\kappa_4$ and then readjusting this coupling according to the measured average volume $N_4$. It is a question of the operator's intuition to distinguish between the usual fluctuation and a systematic mistuning of the coupling. Additionally two frequencies $f_1, f_2$ have to be adjusted to get roughly the same resulting acceptance for each type of move. The frequencies $f_1$ and $f_2$ are numbers between zero and one which say how often to attempt the move 0,4 and 1,3 respectively. The frequency for the move 2 is $1 - f_1 - f_2$. The algorithm for the automatic tuning is depicted in the figure 2. It starts with an initial guess for the coupling $\kappa_4$ and for the frequencies $f_1, f_2$. As a first guess the result of the previous tuning procedure for the closest couplings is used. Then `TRY` updates are executed. From them the algorithm estimates the quality of the current configuration. We use to this purpose the simple quantity $qual = 1 - N_4/N_4^0$. It is compared with two thresholds `sure, good`. When the quality is better than `sure`, $|qual| \leq$ `sure`, the parameter $\kappa_4$ is left unchanged, but the frequencies are readjusted. To prevent from accepting a configuration by accident we check the stability of a configuration. We do this by introducing the counter `SUREC` which sums the number of sequential hits with `sure` quality. Only when `SUREC` is larger than a certain value `secure` a setup is accepted as valid. To reduce the fluctuation of $N_4$, the number of updates `TRY` is set to a larger number `suretry`.

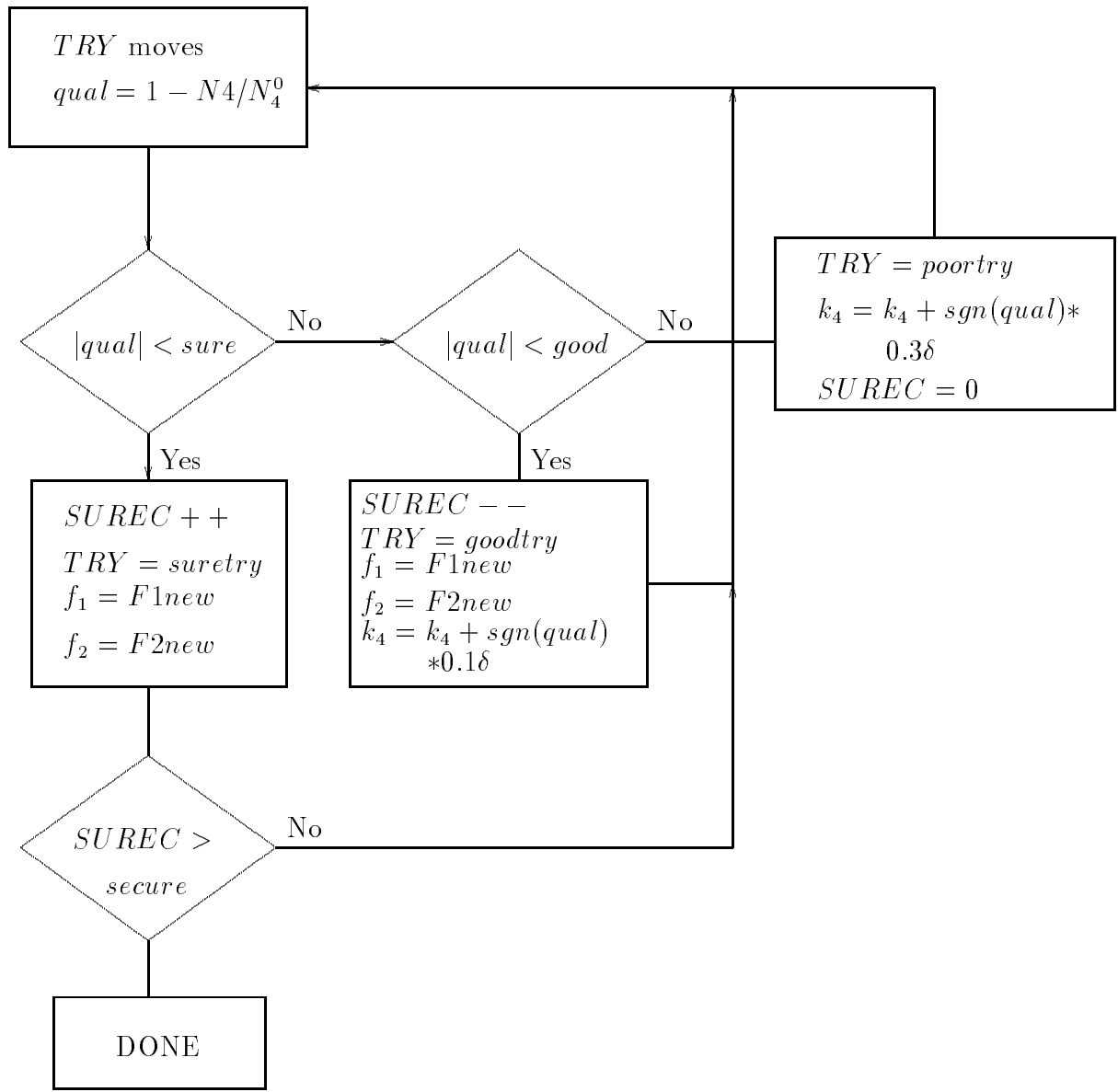

Figure 2: The autotune algorithm

If the quality is worse than `sure` but better than `good` (sure< $|qual|$ ≤ good), the `SECURECounter` is is decreased by one (as far as the result is larger then zero). The coupling $\kappa_4$ is moderately readjusted, as a step size in this region we use $0.1 * \delta$. The number of updates for the next try is set to a value `goodtry` < `suretry`. The frequencies $f_1, f_2$ are also readjusted.

If we are outside a `good`-quality, $\kappa_4$ is tuned in relatively large steps. We use $1/3 * \delta$ as a step size. In this region, the frequencies $f_1, f_2$ are *not* readjusted because if $\kappa_4$ is too far from the correct value $\kappa_4(\beta, \kappa_2)$ the frequencies use to run to unusable values.

## 5. The Program

The elementary updates are coded in the source file MOVE.F. This package supports the geometric moves, update of gauge fields, as well as creating, loading and storing a configuration. This part is machine independent and can be compiled by almost all compilers. The observables are coded in the part MEASURE.F. This package supports the commonly used observables for the gravity sector: the geodesic distances $d_1, d_4$ and the integrated curvature correlation. All measurements are taken after a geometric update when the system reaches for the first time the desired volume $N_4^0$. The main program, GRAVZ2.F, is machine-dependent and works on the PARAGON whose MIMD-architecture allows to run independently simulations for different coupling constants on the compute-nodes of the machine. We performed our simulations on the PARAGON XP/S 10 with up to 64 nodes.

The fine tuning is done either by the interactive program HANDTUNE.F or by AUTOTUNE.F described previously. As an input, AUTOTUNE.F accepts a table :

```
#N4    DN4   Beta    k2       dk4   g:f
4000   500   0.020  -0.100   0.05  1.00
4000   500   0.040  -0.100   0.05  1.00
```

where N4 is the desired volume $N_4^0$, DN4 the maximal fluctuation $\Delta N_4$ from this value, Beta = $\beta$, k2 = $k_2$. dk4 stands for the additional potential $\delta$ used to prevent the system from going too far from $N_4^0$. g:f controls the ratio between frequencies of the geometrical updates compared to the field updates. The output of AUTOTUNE.F

```
#N4    DN4   Beta   k2     k4    dk4  g:f  f1    f2
4000   500   0.020 -0.100  1.374 0.05 1.00 0.066 0.316
4000   500   0.040 -0.100  1.444 0.05 1.00 0.069 0.301
```

is ready to be executed by the production front end GRAVZ2.F when the additional line containing general information about a production run is added at the beginning. The format of this line is

```
fmeas    nmeas   nterm nsave nlog
```

where `fmeas` is the number of updates between measurements, `nmeas` is the number of measurements, `nterm` the number of updates used for thermalization, `nsave` is the configuration save frequency in measurements and `nlog` controls the frequency for a message to `stdout`. GRAVZ2 creates a result-file with a name of the form: r$ss \pm \beta\beta\beta\beta \pm kkkk$. $ss$ is the volume $N4$ expressed in $K$simplices, the beta-field is $\beta * 1000$ and the k-field is $k_2 * 1000$. The configuration file follows the same naming convention where the 'r' is replaced by 'c'. If this file exists, GRAVZ2 restarts this configuration, otherwise it starts from the new configuration built from the minimal sphere. A valid input file for a 2-node job would be:

```
5 10000 250 200 50
4000 500 0.020 -0.100 1.374 0.05 1.00 0.066 0.316
4000 500 0.040 -0.100 1.444 0.05 1.00 0.069 0.301
```

The program produces two result – files `r04+0020-0100`, `r04+0040-0100` and two configuration-files `c04+0020-0100`, `c04+0040-0100`. The format of the results-file is human-readable, as an example we present a few rows from the file `r04+0020-0100`:

```
#!NEWFILE
#!DATE   2-Nov-93  21:43:26
#!STPDSC   n4  dn4 beta k2 k4 dk4 fg f1 f2 mes_fr
#!SETUP 4000 500 0.020 -0.100 1.347 0.05 1. 0.066 0.316 5
#!DTADSC   <D1>  <D4>   N0 <N4> R^2 ssso sss
#!DTABGN
2.718  11.508   603   4008.0 0.1515  0.379  0.082
2.663  11.171   592   4000.0 0.1561  0.401  0.083
2.567  11.205   593   3992.0 0.1555  0.517  0.105
```

The results are presented in [9] and [10].

# Acknowledgements

We are grateful to J. Ambjørn and B. Petersson for discussions and the help in preparing the paper. We want to thank HLRZ–Jülich for the time on Paragon. One of us (ZB) would like to thank Alexander von Humboldt Foundation for the fellowship. This work was partially supported by KBN grant 2P30204705.